# Effect of Impurities in Description of Surface Nanobubbles

Siddhartha Das,* Jacco H. Snoeijer, and Detlef Lohse

*Physics of Fluids, Faculty of Science and Technology, University of Twente, P.O. Box 217, 7500 AE Enschede, The Netherlands*

E-mail: s.das@utwente.nl

∗To whom correspondence should be addressed

**Abstract**

Surface nanobubbles emerging at solid-liquid interfaces of submerged hydrophobic surfaces show extreme stability and very small (gas-side) contact angles. In a recent study Ducker (W. A. Ducker, Langmuir **25**, 8907 (**2009**).) conjectured that these effects may arise from the presence of impurities at the air-water interface of the nanobubbles. In this paper we present a *quantitative* analysis of this hypothesis by estimating the dependence of the contact angle and the Laplace pressure on the fraction of impurity coverage at the liquid-gas interface. We first develop a general analytical framework to estimate the effect of impurities (ionic or non-ionic) in lowering the surface tension of a given air-water interface. We then employ this model to show that the (gas-side) contact angle and the Laplace pressure across the nanobubbles indeed decrease considerably with an increase in the fractional coverage of the impurities, though still not sufficiently small to account for the observed surface nanobubble stability. The proposed model also suggests the dependencies of the Laplace pressure and the contact angle on the type of impurity.

# I. INTRODUCTION

Over many years researchers have been fascinated by a number intriguing, yet not well understood, phenomena that occur when water comes in contact with a hydrophobic (non-wetting) substrate. The presence of the hydrophobic surface leads to the formation of spherical cap-like bubbles at the solid-liquid interface, called "surface nanobubbles". Over the years AFM techniques have been the most popular method in studying these surface nanobubbles [1-5]. Depending on the conditions that lead to their formation, different behaviors of the nanobubbles have been found by these studies: e.g., their spherical cap-like shape and chances of deviation from that shape [6-8], merging of two adjacently located nanobubbles [6,9], disappearance of nanobubbles in case the water is degassed [10], possible reappearances by exchange of solvents [7,11-15] or increase of temperature [11], or electrolysis [9,16] etc. The different relevant issues pertaining to the formation and behavior of surface nanobubbles are well summarized in a very recent review by Hampton and Nguyen [17].

The challenges concerning the surface nanobubbles stem from the fact that, unlike the macroscopic or even microscopic bubbles, one cannot explain their properties at equilibrium using the known standard values of surface tension for the media involved. Two of the biggest mysteries concerning nanobubbles are their extremely small (gas-side) contact angle ($\theta$; see figure 1) and their extremely large stability. For example, for bubbles at octadecyltrichlorosilane(OTS)-silicon-water interface (where $\Delta\sigma = \sigma_{sl} - \sigma_{sg} = 0.025 N/m$ and $\sigma_{lg} = 0.072 N/m$) [18], by employing Young's equation (established for macroscopic situations) for the contact angle $\theta$ (figure 1) one gets:

$$\theta = \cos^{-1}\left(\frac{\Delta\sigma}{\sigma_{lg}}\right) = 70^0. \tag{1}$$

This indeed is the contact angle of the macroscopic bubbles on that material. However, in case of nanobubbles, different experiments reveal much smaller values ($20^0 - 30^0$) of the gas-side contact angle at the OTS-silicon-water interface [7,18,19].

The other main mystery around surface nanobubbles is their extreme stability. Investigations report nanobubbles to remain stable for over days when left undisturbed [12,15]. Simple calculations of the Laplace pressure ($\Delta p$) for a nanobubble of radius $R_b$=100 nm (at the OTS-silicon-water interface, for example, with $\sigma_{lg}$ = 0.072$N/m$), however, gives an estimate of

$$\Delta p = \frac{2\sigma_{lg}}{R_b} = 1.44 MPa, \quad (2)$$

for the Laplace pressure, suggesting that a nanobubble would dissolve in milliseconds in case that macroscopically established laws are applied [20].

Various explanations have been proposed to resolve these two mysteries. Effects like negative line tension [21] or pseudo partial wetting [6] have been argued to be responsible for the unexpectedly small (gas-side) contact angle. On the other hand, the issue of superstability has been addressed by postulating a compensating gas influx into the bubble at the contact line [20], thanks to the attraction of gas molecules towards the hydrophobic walls [22]. This influx then balances the gas outflux from the nanobubble, leading to bubble stability. Other explanations include possible lowering of surface tension for large curvatures on small scales [23,24], the oversaturation of liquid with gas in the vicinity of nanobubbles [15], the effect of induced charges in the Debye layer developed around the bubble interface [25], etc. A recent study by Borkent et al. [8] suggests that contaminations also have a strong effect on the contact angle. As pointed out in a recent study by Ducker [18], the presence of such impurities can act as a shield to the outflux of gases (making the bubbles more stable) and at the same time can lower the effective value of the liquid-gas surface tension $\sigma_{lg}$. Smaller $\sigma_{lg}$ can indeed potentially explain the small contact angle (see eq. (1)) and the large stability (caused by smaller values of $\Delta p$, see eq. (2)) of the nanobubbles.

In this study, we propose a theoretical framework to quantitatively investigate to what extent the presence of impurities can affect the contact angle and the stability of the nanobubbles. We only investigate the consequence of lowering of $\sigma_{lg}$ with the impurities, and not the prevention of outflux of the gases [18]. The model is based on the equilibrium description of surfactant adsorption [26] and can be employed to quantify the impurity/surfactant-induced lowering of surface tension of any general air-water interface. To test the generality of this model, we reproduce with this model (with realistic fitting parameters) the classical experimental results of surfactant-induced lowering of surface tension for both ionic [27,28] and non-ionic [29,30] surfactants. This general model is next applied to the nanobubble-impurity system.

The adsorption time scale for most of the known surfactants (we treat the impurities as surfactants) is of the order of few seconds and consequently the surface tension attains the reduced value in this time [31-33]. On the contrary, the typical time scale that can be ascribed to the formation or morphological changes (if any) of the surface nanobubbles is at least of the order of 10 minutes [5,34,35]. Hence, except for the initial transients that last for very small time, the surface nanobubbles are expected to be formed in presence of the constant reduced value of the air-water surface tension, thereby allowing us to invoke equilibrium treatment. In this model the chemical potential of the impurities (and the solvent) in the bulk is identical to that at the air-water interface (or the surface layer) of the nanobubbles. The analysis will be performed for both non-ionic as well as ionic impurities. The results for these cases are obtained as functions of the degree of surface coverage of the impurities. We thus start from an equilibrium picture that says that a given amount of impurity is already present at the surface layer, without trying to resolve the possible mechanism of such impurity adsorption at the surface layer. The proposed model can be used to investigate the effects of different factors pertaining to the impurities (e.g., their size, number of types of impurities, their nature i.e., ionic or non-ionic) on the contact angle $\theta$ and the Laplace pressure $\Delta p$ of the nanobubbles. Considering the case of nanobubbles formed at the OTS-silicon-water interface as an example, it is shown that the impurities indeed lower the contact angle $\theta$, and for a significantly high surface coverage of impurities, the value predicted is quite close to those found by experiments [7,19]. In fact, the nanobubble contact angles are found to be in the range that are explained by experimental evidence of adsorption of common surfactants to air-water interface [36,37]. Impurities, in sufficient concentration, can also significantly reduce the Laplace pressure $\Delta p$ (approximately to half the value predicted by eq. (2)). This value of Laplace pressure, however, is still large enough to enforce extremely fast diffusion of gases from the nanobubbles rendering it unstable. In fact, the equilibrium adsorption of soluble surfactants to water (the picture which is quantified in this paper) never, in practice, reduces the surface tension of water below 0.025-0.03 N/m, which is still not small enough to account for the observed surface nanobubble stability. Thus, we have not solved the puzzle of nanobubble "superstability". However, we believe there is no one single effect that makes the nanobubble so stable. Rather, nanobubble superstability is a result of a number of different factors acting simultaneously and the presence of impurities at the air-water interface can indeed be considered as one of them. In addition, as suggested by Ducker [18], there is also a possibility that some insoluble impurity molecules may get stuck on the bubble, decreasing the surface tension much below 0.025-0.03 N/m thereby ensuring that the Laplace pressure becomes small enough to enforce nanobubble superstability.

## II. THEORY

Liquid-vapor surface tension is a manifestation of the strength of the molecular attractive forces experienced by the layer of solvent molecules present at the interface. In cases of solvents like water where the molecules can form hydrogen bonds (HB), the surface tension is the result of the dispersion interaction forces (present for all types of solvents) and the HB induced interaction forces (i.e., $\sigma_{lg} = \sigma_{lg}^{HB} + \sigma_{lg}^{d}$) [36]. However, the contribution of the HB effect is much larger than the dispersion effect, and consequently the surface tension for water is much higher (~0.072 N/m) as compared to other solvents which do not form HB, e.g., oil (liquid-vapor surface tension ~0.025 N/m). When impurities are present at the liquid-gas interface of nanobubbles, a water molecule at the interface gets surrounded by the impurity molecules preventing it to successfully hydrogen-bond with neighboring water molecules at the interface. This significantly lowers the HB induced attractive forces, thereby considerably lowering the surface tension. The lowering of the surface tension is defined as the surface pressure $\Pi (= \sigma_{lg} - \sigma_{lg}^{/}$; where $\sigma_{lg}^{/}$ is the reduced surface tension due to impurity effect). As the dispersion effects are universally present, irrespective of the extent of the fractional coverage of the impurities, the surface tension must always contain the contribution of the dispersion effects. This will mean $(\Pi)^{max} = (\sigma_{lg})^{max} - (\sigma_{lg}^{/})^{min} = (\sigma_{lg}^{HB} + \sigma_{lg}^{d}) - (\sigma_{lg}^{d}) = \sigma_{lg}^{HB}$ (based on the assumption that the effect of impurities, at most, can completely block out the contribution of the HB interaction on surface tension, so that $(\sigma_{lg}^{/})^{min} = \sigma_{lg}^{d}$). As all the results presented below consider only the effect of this difference of surface tensions (induced by disregarding the HB-interaction by the impurities), there will be a maximum value of surface coverage of impurities (this maximum value varies from case to case) at which $\Pi = \Pi^{max}$ and beyond that value of surface coverage the surface pressure no longer changes (i.e., becomes constant at $\Pi^{max}$) with increase in the concentration of impurities. Thus this maximum surface coverage of impurities is analogous to the role played by the critical micelle concentration (cmc) in studies delineating the effects of surfactants on surface tension, where for concentration above cmc, there is no further change in surface tension with increase in concentration [38-40]. Note that once this critical concentration is reached, with further increased concentration, there are chances that (under pressure fluctuations) the nanobubble may actually buckle, see Marmottant et al. [41], reducing the surface tension to effectively zero. Such a situation would demand an analysis far beyond the simple analysis of eq. (2) and is beyond the scope of the present paper. Thus in the results to be presented below, it is implied that for a given case (non-ionic or ionic impurities) we always operate at a concentration (of impurities) regime which is less than this critical concentration value (beyond this value the surface pressure is constant) of impurities.

We start our analysis by considering the equilibrium condition of type $i$ impurity adsorbed at the air-water interface of the surface nanobubbles. At equilibrium, the chemical potential of the impurity in the surface layer (liquid-gas interface) must be equal to its chemical potential in the bulk. This allows us to invoke the Butler equation [26,42] describing the chemical equilibrium of the impurity of type $i$, so that one can write

$$\mu_i^{0s} - \omega_i \sigma_{lg}^{/} + RT \ln\left(f_i^s x_i^s\right) = \mu_i^{0b} + RT \ln\left(f_i^b x_i^b\right). \tag{3}$$

In eq. (3), $\mu_i^{0s}$ is the standard state chemical potential of impurity $i$ in the surface layer, and $f_i^s$ and $x_i^s$ are the activity coefficient and the mole fraction of impurity $i$ in the surface layer, respectively. Here the subscript "s" refers to the surface layer (air-water interface). Similarly, $\mu_i^{0b}$ is the standard state chemical potential of impurity $i$ in the bulk solution and $f_i^b$ and $x_i^b$ are the activity coefficients and the mole fraction of impurity $i$ in the bulk solution, respectively. The subscript "b" refers to the bulk solution. $\omega_i$ is the partial molar area of the impurity $i$ in the surface layer and $\sigma_{lg}^{/}$ is the modified value of the surface tension in presence of impurities. In eq. (3), the second term on the LHS is the contribution due to the surface coverage of the air-water interface (by the impurity of type $i$). Finally, the last terms on LHS and RHS represent the effect of mixing at the surface layer and the bulk, respectively.

Using the same Butler equation, one can similarly write the equilibrium equation for the solvent (denoted by subscript $i = 0$) as

$$\mu_0^{0s} - \omega_0 \sigma'_{lg} + RT \ln\left(f_0^s x_0^s\right) = \mu_0^{0b} + RT \ln\left(f_0^b x_0^b\right). \tag{4}$$

To evaluate $\sigma'_{lg}$ from eqs. (3) and (4) in the presence of impurities, it is necessary to first connect the values of the chemical potentials of the solvent and the impurities at the standard state (at the surface layer and the bulk solution) with the surface tension $\sigma_{lg}$ ($\sigma_{lg}$ is the surface tension without the effect of impurities). For the solvent, one can write for the standard state

$$x_0^s = 1, f_0^s = 1, x_0^b = 1, f_0^b = 1. \tag{5}$$

From eqs. (4) and (5), we then obtain

$$\mu_0^{0s} - \omega_0 \sigma_{lg} = \mu_0^{0b}. \tag{6}$$

For the $i^{th}$ impurity (or surface active) component, the standard state will imply an infinitely dilute solution. This means

$$x_i^b \to 0, f_i^b = f_{(0)i}^b, f_i^s = f_{(0)i}^s, \sigma'_{lg} = \sigma_{lg}. \tag{7}$$

In eq. (7), we denote the conditions at infinite dilution by the additional subscript "(0)".

Using eqs. (3) and (7), we get

$$\mu_i^{0b} - \mu_i^{0s} = -\omega_i \sigma_{lg} + RT \ln(K_i) + RT \ln\left(\frac{f_{(0)i}^s}{f_{(0)i}^b}\right), \tag{8}$$

where $K_i = \left(x_i^s / x_i^b\right)_{x_i^b \to 0}$.

Next, using eqs. (4) and (6), we obtain the equation of state of the solvent at the surface layer as

$$\ln\left(\frac{f_0^s x_0^s}{f_0^b x_0^b}\right) = -\frac{\omega_0 \left(\sigma_{lg} - \sigma'_{lg}\right)}{RT} = -\frac{\omega_0 \Pi}{RT}, \tag{9}$$

where $\Pi = \sigma_{lg} - \sigma'_{lg}$ is the surface pressure or the extent of lowering of the surface tension due to presence of impurities.

Similarly, using eqs (3) and (8), we obtain the equation of state of the impurity of type $i$ at the surface layer as

$$\ln\left(\frac{f_i^s x_i^s / f_{(0)i}^s}{K_i f_i^b x_i^b / f_{(0)i}^b}\right) = -\frac{\omega_i \left(\sigma_{lg} - \sigma'_{lg}\right)}{RT} = -\frac{\omega_i \Pi}{RT}. \tag{10}$$

We can simplify the equations of state of the solvent and the impurities further by assuming ideality of the bulk solution (i.e., zero enthalpy or entropy of mixing). For the present case the bulk concentration of the impurities is assumed to be relatively small, implying

$$f_0^b = 1, x_0^b = 1, f_{(0)i}^b = 1, f_i^b = 1 \tag{11}$$

Using eq. (11) in eqs. (9) and (10), we can obtain simplified expressions for the equation of state of the solvent and the impurities in the surface layer as

$$\ln(f_0^s x_0^s) = -\frac{\omega_0 \Pi}{RT}, \tag{12}$$

and

$$\ln\left(\frac{f_i^s x_i^s}{K_i f_{(0)i}^s x_i^b}\right) = -\frac{\omega_i \Pi}{RT}. \tag{13}$$

Eq. (12) can be expressed in terms of the mole fraction of the impurities as

$$\Pi = -\frac{RT}{\omega_0}\left[\ln\left(1 - \sum_{i \geq 1} x_i^s\right) + \ln(f_0^s)\right]. \tag{14}$$

For a general case of *n* types of impurities, eqs. (13) and (14) represent a system of *n* + 1 equations in *n* + 1 unknowns (namely $x_1^s, x_2^s, \ldots x_n^s$ and $\Pi$). However, to obtain a complete solution of these *n* + 1 equations, one needs to know a large number of parameters beforehand, namely, $\omega_0, f_0^s$ (parameters pertaining to the solvent) and $\omega_1, \omega_2, \ldots, \omega_n, f_{(0)1}^s, f_{(0)2}^s, \ldots f_{(0)n}^s, f_1^s, f_2^s, \ldots f_n^s, K_1, K_2, \ldots, K_n$ (parameters pertaining to the impurities).

In the present study, we will not solve for such a general situation; rather we will take up two simple cases that may successfully portray the effect of presence of impurities in altering the surface tension and the nanobubble parameters. The first one is the case of non-ionic impurities, whereas the second one is that of ionic impurities. For both of these cases, we simplify the situation assuming that the surface layer is ideal, which will mean that the activity coefficient of all components in the surface layer is equal to unity, i.e.

$$f_{(0)1}^s = f_{(0)2}^s = \ldots = f_{(0)n}^s = f_0^s = f_1^s = \ldots = f_n^s = 1. \tag{15}$$

Such an assumption on ideality of the surface layer can have possible limitations for cases with large fractional coverage of impurities (leading to very large surface pressures), where the impurities may interact with each other necessitating the use of non-ideal surface layer condition [43,44]. Effects of such non-idealities will be discussed in a future paper, but for the present study we restrict our treatment to an ideal surface layer. We further assume that the values of partial molar areas of the impurities are identical i.e., ($\omega_1 = \omega_2 = .... = \omega_n$). Under this condition, the mole fraction of the impurities are identical to their respective fraction coverage $\beta_i$ [26] (where $\beta_i = \omega_i \Gamma_i$, where $\Gamma_i$ is the adsorption of the impurity of type $i$). In presence of these simplifying conditions, we try to obtain the partial pressure for the non-ionic and ionic impurities.

### A. Non-ionic impurities

Using eq. (15), along with the condition of identical partial molar areas of the impurities, eqs. (13) and (14) get simplified to

$$\beta_i = K_i x_i^b \exp\left(-\frac{\omega_i \Pi}{RT}\right), \tag{16}$$

and

$$\Pi = -\frac{RT}{\omega_0}\left[\ln\left(1 - \sum_{i \geq 1} \beta_i\right)\right]. \tag{17}$$

Interestingly, eq. (17) establishes that in order to obtain the surface pressure, it is sufficient to know the total fractional coverage of the impurities at the surface layer, without requiring the value of coverage of individual types of impurities (which, if required, can be obtained by using eq. (16)). Once the surface pressure is known, one can calculate the modified value of the surface tension $\sigma'_{lg}$ as

$$\sigma'_{lg} = \sigma_{lg} - \Pi. \tag{18}$$

This modified value of the surface tension is next used to replace $\sigma_{lg}$ to calculate the modified values of the contact angle and the Laplace pressure using eqs. (1) and (2).

### B. Ionic impurities

For cases where ionic impurities are adsorbed at the surface layer, their mutual repulsion will result in an additional surface pressure $\Delta\Pi_{ionic}$. This contribution to surface pressure act in addition to the contribution due to surface coverage (discussed previously). To calculate $\Delta\Pi_{ionic}$ it is first considered that the presence of ionic impurities creates a Dielectric Double Layer (DEL) at the interface, leading to a charge separation between the ionic Double Layer and the neutral bulk solution. One can subsequently invoke Gouy-Chapman theory to calculate the energy of this double layer system [45-48]. This energy is provided by the original surface energy of the interface (i.e., without impurities), and consequently the surface energy decreases, lowering the surface tension. This lowering, equal to the surface pressure $\Delta\Pi_{ionic}$, is thus the per unit area energy of the double layer system. Consequently, one can write (under the assumption that at the surface there is only one kind of ionic impurity) [45-48]

$$\Delta\Pi_{ionic} = \frac{4RT}{F}\sqrt{2.10^3 \varepsilon_0 \varepsilon_r RT c_\Sigma}\left[\cosh\left(\frac{z_s F \psi_s}{2RT}\right) - 1\right]. \qquad (19)$$

In eq. (19) $F$ is the Faraday constant, $\varepsilon_0$ is the permittivity of the vacuum, $\varepsilon_r$ is the relative permittivity of the medium, $c_\Sigma$ is the total bulk ionic concentration (in moles/litre), $z_s$ is the valence of the impurity ions at the surface layer and $\psi_s$ is the electrical potenial of the surface layer.

The surface potential can be obtained as a function of the adsorption value ($\Gamma_s$) of the ionic impurity species (the extent of the adsorption dictates the surface charge density of the layer, assuming that all the charged species adsorbed at the surface act as surface active ions, i.e., are those ions that define the potential) as [45-48]

$$\sinh\left(\frac{z_s F \psi_s}{2RT}\right) = \frac{F|z_s|\Gamma_s}{\sqrt{8.10^3 \varepsilon_0 \varepsilon_r RT c_\Sigma}}. \qquad (20)$$

Using $\varepsilon_r = 79.8$ for water and $T = 300K$, we get

$$\sinh\left(\frac{z_s F \psi_s}{2RT}\right) = \frac{8.15 \times 10^5 |z_s|\Gamma_s}{\sqrt{c_\Sigma}} = \frac{8.15 \times 10^5 |z_s|\beta_1}{\omega_1 \sqrt{c_\Sigma}}, \qquad (21)$$

using $\Gamma_s = \beta_1/\omega_1$, where $\beta_1$ and $\omega_1$ are the fractional coverage and the partial molar area of the ionic impurity at the interface, respectively.

Now for the case with relatively small bulk concentration (in moles/liter) of the ionic impurities and $\omega_1 \sim O(10^5 - 10^6\ m^2/mol)$, we will always have

$$\sinh\left(\frac{z_s F \psi_s}{2RT}\right) \gg 1. \qquad (22)$$

As $z_s$ and $\psi_s$ are always of identical sign, the argument $\frac{z_s F \psi_s}{2RT}$ is always positive. Again, when $\sinh(y) \gg 1$ and $y \gg 0$, we can safely approximate $\sinh(y) \approx \cosh(y)$, which will mean

$$\sinh\left(\frac{z_s F \psi_s}{2RT}\right) \approx \cosh\left(\frac{z_s F \psi_s}{2RT}\right) \approx \exp\left(\frac{z_s F \psi_s}{2RT}\right). \tag{23}$$

Using eqs. (19), (20) and (24), we get [49,50]

$$\Delta \Pi_{ionic} \approx 2RT|z_s|\Gamma_s = \frac{2RT|z_s|\beta_1}{\omega_1}. \tag{24}$$

Consequently, the expression of the surface pressure (under the approximation that there is only one kind of impurity at the interface and that impurity is ionic in nature, with all the ions acting as surface active or potential-determining ions) becomes

$$\Pi = -\frac{RT}{\omega_0}\left[\ln(1-\beta_1)\right] + \frac{2RT|z_s|\beta_1}{\omega_1}. \tag{25}$$

With this modified value of the surface pressure (incorporating the effects of ionic impurities), one can employ eqs. (18), (1) and (2) to obtain $\sigma_{lg}^{/}$, $\theta$ and $\Delta p$ for the case with ionic impurities.

## III. RESULTS AND DISCUSSIONS

In this section we shall first provide the experimental validation of the general mathematical framework developed in the previous section. Next, we shall extend this model to quantify and discuss the effects of surface impurities in altering the nanobubble parameters for the two cases described in the previous section. Finally we will analyze the importance of the present study in the light of the existing experimental evidences and suggest a possible experiment that may be performed to validate the proposed theory, in context of the surface nanobubbles.

### A. Experimental validation of the effect of impurities on surface tension

The relationship between the surface pressure and the surface coverage of impurities are illustrated through eqs. (17) and (25). Eqs. (17) and (25) are applicable to any general air-water interface in presence of impurities (treated as surfactant molecules). Thus with the choice of the correct parameters, these equations can be successfully employed to validate the experimental observations of the surfactant-induced lowering of surface tension at the air-water interfaces. The experimental results, however, invariably predicts the surface pressure as a function of the bulk concentration (and not surface coverage) of surfactants (ionic or non-ionic). In the present model, to obtain the surface pressure as a function of the bulk concentration, eq. (16) is considered in addition to eqs. (17) and (25). For non-ionic surfactants, under the condition that only one kind of surfactant is present, eq. (16) and eq. (17) are iteratively solved to obtain the surface pressure as a function of the bulk concentration $c_b$ ($c_b$ being expressed in moles/m$^3$, it can be related to the bulk mole fraction $x_b$ as $x_b = \dfrac{c_b}{c_b + c_b^w}$, where $c_b^w$ is the number of moles of water in a volume of 1 m$^3$, i.e., $c_b^w = \dfrac{1000}{0.018} = 5.556 \times 10^4$ moles ). For the non-ionic surfactants, the results from the present simulation are validated with experimental results for surfactant BHBC$_{16}$ [29,30] (See figure 2a). For the present model the following parameters are considered: $\omega_0 = 6.023 \times 10^4$ m$^2$/mol, $\omega_1 = 2.5 \times 10^5$ m$^2$/mol [26] and $K_1$ (used as a fitting variable) = $5.9 \times 10^7$. Typically the parameter $K_1$ is around 1 order or even higher than the magnitude of parameter $b_1$, classically defined for surfactants adsorption (from Ref. 26, one can write $K_1 = b_1 c_1 / x_1 = 55.56 \times b_1$, with 55.56 representing the number moles of water in a volume of 1 liter). For BHBC$_{16}$, $b = 1.6$ $l$/mol [26], which justifies the order of magnitude for the above choice of $K_1$. For the ionic impurities, results can be obtained by iteratively solving eqs. (16) and (25). For this case the results are validated with experimental results for the case with surfactant Sodium Dodecyl Sulfate (SDS) with no added inorganic salt [27,28] (See figure 2b). In this case the parameters are: $\omega_0 = 6.023 \times 10^4$ m$^2$/mol, $\omega_1 = 3.4 \times 10^5$ m$^2$/mol (estimated from the SDS partial molar volume value of $2.6 \times 10^{-4}$ m$^3$/mol [51]) and $K_1$ (used as a fitting variable) = $4.0 \times 10^3$ (this choice is justified by the data for the corresponding $b_1$ [52] as 39.1 $l$/mol). There is excellent match of the prediction from the present simulation with the experimental results for non-ionic surfactants. Also for the ionic surfactants the match is good except for very low concentration; although it must be mentioned here that by an approximate extrapolation of the experimental data, we may get a surprising result of zero surface pressure for finite ionic concentration. In summary, we can infer that our simulation results can pretty well match the experimental data and hence we conclude that the proposed model can indeed be used for calculating the surfactant-induced lowering of any general air-water interface. In the following sections, we apply this theory to obtain the reduced surface tension and the resulting changes in the contact angle and the Laplace pressure for the surface nanobubbles.

### B. Effect of non-ionic impurities on surface nanobubbles

Figures 3a-c depict the variation of the modified surface tension ($\sigma_{lg}^{/}$), the nanobubble gas-side contact angle θ and the Laplace pressure Δ$p$, respectively, with the fractional coverage of impurities for different possible values of the partial molar surface area of the water ($\omega_0$) (for nanobubbles formed at the OTS-silicon-water interface). Here the plots are provided as functions of the fractional coverage of impurities and not the impurity bulk concentration, so as to avoid the use of fitting constant *K*. Theoretical estimates of the size of the water molecules suggest a value of $\omega_{0,th}$ = 6.023×10$^4$ *m$^2$/mol* or (0.1*nm$^2$/molecule*) [26]. However, experimental data suggest some deviation (to higher values) from this theoretical value of $\omega_0$ [26]. As we do not know the exact value of $\omega_0$ to be used, we will present the results for several $\omega_0$ values (including $\omega_0 = \omega_{0,th}$); though the gross order of magnitude remains virtually the same. Figures 3a-c portray that the increase in fractional coverage of the impurities as well as smaller $\omega_0$ values lower the modified surface tension, leading to a smaller gas-side contact angle as well a smaller Laplace pressure. For significantly large fractional coverage, the gas side contact angle is approximately close to the one predicted by experimental findings (for nanobubbles at OTS-silicon-water interface) [7,19]. However, for the Laplace pressure, even with significant fractional coverage the value remains much higher than the atmospheric pressure. This implies that the phenomenon of superstability of nanobubbles [12] could not be explained by the effect of soluble impurities alone; rather this effect could be looked upon as one of the possibly many factors that are simultaneously operative in ensuring the large stability of the nanobubbles. Lowering of surface tension (and the resulting changes of the nanobubble parameters) with a decrease of $\omega_0$ can also be interpreted from a more physical perspective. Larger $\omega_0$ values indicate that the effective space occupied by a water molecule in the surface layer is large, which means that there is a greater chance that due to steric effects the water molecules remain preferably less surrounded by the impurity molecules and more surrounded by neighboring water molecules, allowing it to form HB with them. Consequently, a smaller $\omega_0$ leads to larger lowering of the HB interaction effect, leading to a larger surface pressure and more pronounced lowering of θ and Δ$p$.

### C. Effect of ionic impurities on surface nanobubbles

The extent to which the ionic nature of the impurities can change the values of the variables like $\sigma_{lg}^{/}$ and the nanobubble parameters θ and Δ$p$ are illustrated in figures 4a-c, which plot these quantities as function of the fractional coverage of impurities with and without considering ΔΠ$_{ionic}$. The cases with ΔΠ$_{ionic}$ are plotted for different values of the partial molar area of the impurities $\omega_1$. To obtain these plots it is assumed that there is only one kind of impurity at the interface and that impurity is ionic in nature, with all the ions acting as surface active or potential-determining ions. It is clearly exhibited, as has been argued in section II B, that in case the impurities become ionic, the effect of impurities becomes even more pronounced in affecting the nanobubble parameters (provided all other things remain identical). For example, the extent of lowering of θ and Δ$p$ that are achieved with a fractional coverage of 0.6 for non-ionic impurities (See in figures 3b,c plots corresponding to $\omega_0 = \omega_{0,th}$), are now exhibited for a fractional coverage of 0.5 for ionic impurities (with $\omega_1 = 4\times10^5$ $m^2/mol$). Physically, this points to the fact that to ensure that a given number of similarly charged ions are allowed to remain adsorbed simultaneously at the interface there needs to be significant expenditure of the original surface energy of the interface. In case the impurity ions are smaller in sizes (characterized by smaller values of $\omega_1$), there are larger number of impurity ions (for a given value of fractional surface coverage), which will mean that the total number of repelling electrostatic interactions (between these similarly charged ions at the interface) increases, requiring an even larger expenditure of the original surface energy to keep them at the surface layer. Hence ΔΠ$_{ionic}$ becomes higher for smaller $\omega_1$, leading to more pronounced lowering of $\sigma_{lg}^{/}$, θ, and Δ$p$ (see figures 4a-c).

### D. Usefulness of the proposed theory and its possible experimental verification in context of surface nanobubbles

Although the quantification of the effect of surface impurities on nanobubble equilibrium properties has hitherto hardly been available in the literature, there have been experimental evidences and qualitative explanations on the possible impact of the presence of impurities at the air-water interface of the surface nanobubbles [8,18]. These studies establish that a number of apparently non-intuitive characteristics of the surface nanobubbles originate from the presence of impurities at the air-water interface. As pointed out in a recent study by Borkent et al. [8], possible contaminants can be siloxane oil and other polymeric organic derivatives of high molecular weight silicon compounds like PDMS. The primary source of these contaminants are the AFM cantilevers used to detect the nanobubbles [8], as well as the substrates where the nanobubbles are formed. As the exact nature of the contaminants are not yet clearly known, such a general mathematical framework proposed in this study to describe the surface nanobubble parameters as a function of the nature (ionic or non-ionic) of the impurities is extremely useful for the purpose of sketching a comprehensive quantitative picture.

One can suggest a direct experimental procedure to verify the proposed theory. The system with surface nanobubbles needs to be subjected to both static and dynamic light scattering. Depending on whether the system is clean or contaminated, the extent of scattering will be vastly different. This is based on the principle, as suggested in a recent paper by Ducker and his coworkers [53], that it is primarily the impurities, and not the nanobubbles, which cause the scattering. From the scattering measurements one can accordingly quantify the concentration of the impurities in the bulk and at the nanobubble air-water interface and hence attempt to validate the viability of eq. (14), relating the surface pressure (or in effect the nanobubble parameters) with the fraction coverage of impurities.

## IV. CONCLUSIONS

In this paper we develop a general mathematical framework based on equilibrium description of surfactant adsorption at air-water interfaces to analyze the effects of surfactants/impurities in lowering the overall surface tension. This model is subsequently used to study the recently conjectured hypothesis by Ducker [18] that the unexpectedly small (gas-side) contact angle and extremely large stability of surface nanobubbles (created at the solid-liquid interface of submerged hydrophobic surfaces) can partly be explained by possible presence of impurities at the air-water interface of the nanobubbles. Results demonstrate that for significantly high surface coverage of impurities, the (gas-side) contact angle can significantly reduce and indeed exhibit a value close to that suggested by experimental findings [7,19]. Such lowering of the contact angle is similar to that which are suggested by experimental evidences of equilibrium adsorption of common surfactants to water [36,37]. The Laplace pressure ($\Delta p$) is also reduced due to impurity effect, although it still remains large enough to forbid stability of nanobubbles. The finding that the equilibrium adsorption of soluble surfactants to water can only reduce the surface tension to around 0.025-0.03 N/m ensuring that the Laplace pressure is still rather high implies that the equilibrium surfactant-adsorption model (presented in this paper) can only explain the very small gas-side contact angle, but not the long-term stability of surface nanobubble. The analysis is performed for both non-ionic and ionic impurities in an ideal surface layer. With all other parameters remaining identical, for the ionic case, the effect of impurities is found to get even more magnified, dictated by the partial molar area of the impurity molecules. In future studies, we intend to show that the mystery of nanobubble superstability may be further enlightened by accounting for the *non-ideality effects* at the nanobubble air-water interface as well as considering the *disjoining pressure interactions* arising from the possible self-attributed (i.e., without any external contaminant) charged nature of the nanobubble air-water interface.

# References

[1] S. Lou, Z. Ouyang, Y. Zhang, X. Li, J. Hu, M. Li, and F. Yang, J. Vac. Sci. Technol. B **18**, 2573 (2000).

[2] N. Ishida, T. Inoue, M. Miyahara, and K. Higashitani, Langmuir **16**, 6377 (2000).

[3] G. E. Yakubov, H.-J. Butt, and O. I. Vinogradova, J. Phys. Chem. B **104**, 3407 (2000).

[4] A. Carambassis, L. C. Jonker, P. Attard, and M. W. Rutland, Phys. Rev. Lett. **80**, 5357 (1998).

[5] J. W. G. Tyrrell and P. Attard, Phys. Rev. Lett. **87**, 176104 (2001).

[6] A. C. Simonsen, P. L. Hansen, and B. Klosgen, J. Coll. Int. Sci. **273**, 291 (2004).

[7] X. H. Zhang, N. Maeda, and V. S. J. Craig, Langmuir **22**, 5025 (2006).

[8] B. M. Borkent, S. S. de Beer, F. Mugele, and D. Lohse, Langmuir **26**, 260 (2010).

[9] L. Zhang, Y. Zhang, X. Zhang, Z. Li, G. Shen, M. Ye, C. Fan, H. Fang, and J. Hu, Langmuir **22**, 8109 (2006).

[10] X. H. Zhang, G. Li, N. Maeda, and J. Hu, Langmuir **22**, 9238 (2006).

[11] S. Yang, S. M. Dammer, N. Bremond, H. J. W. Zandvliet, E. S. Kooij, and D. Lohse, Langmuir **23**, 7072 (2007).

[12] B. M. Borkent, S. M. Dammer, H. Schonherr, G. J. Vancso, and D. Lohse, Phys. Rev. Lett. **98**, 204502 (2007).

[13] X. H. Zhang, A. Khan, and W. A. Ducker, Phys. Rev. Lett. **98**, 136101 (2007).

[14] S. Yang, E. S. Kooij, B. Poelsema, D. Lohse, and H. J. W. Zandvliet, Euro. Phys. Lett. **81**, 64006 (2008).

[15] X. H. Zhang, A. Quinn, and W. A. Ducker, Langmuir **24**, 4756 (2008).

[16] S. Yang, P. Tsai, E. S. Kooij, A. Prosperetti, H. J. W. Zandvliet, and D. Lohse, Langmuir **25**, 1466 (2009).

[17] M. A. Hampton and N. V. Nguyen, Adv. Coll. Int. Sci. **154**, 30 (2010).

[18] W. A. Ducker, Langmuir **25**, 8907 (2009).

[19] N. Ishida, T. Inoue, M. Miyahara, and K. Higashitani, Langmuir **16**, 6377 (2000).

[20] M. P. Brenner and D. Lohse, Phys. Rev. Lett. **101**, 214505 (2008).

[21] N. Kameda and S. Nakabayashi, J. Coll. Int. Sci. **461**, 122 (2008).

[22] S. M. Dammer and D. Lohse, Phys. Rev. Lett. **96**, 206101 (2006).

[23] C. Fradin *et al.*, Nature **403**, 871 (2000).

[24] S. Mora *et al.*, Phys. Rev. Lett. **90**, 216101 (2003).

[25] F. Jin, J. Li, X. Ye, and C. Wu, J. Phys. Chem. B **111**, 11745 (2007).

[26] V. B. Fainerman, E. H. Lucassen-Reynders, and R. Miller, Coll. Surf. A **143**, 141 (1998).

[27] K. J. Mysels, Langmuir **2**, 423 (1986).

[28] P. H. Elworthy and K. J. Mysels, J. Coll. Int. Sci. **21** 331 (1966).

[29] R. Wustneck, R. Miller, J. Kriwanek, and H.-R. Holzbauer, Langmuir **10**, 3738 (1994).

[30] V. B. Fainerman, R. Miller, R. Wustneck, and A. V. Makievski, J. Phys. Chem. **100**, 3054 (1995).

[31] J. Eastoe and J. S. Dalton, Adv. Coll. Int. Sci. **85**, 103 (2000).

[32] A. J. Prosser and E. I. Franses, Coll. Surf. A **178**, 40 (2001).

[33] J. Liu, C. Wang, and U. Messow, Coll. Polym. Sci. **283**, 139 (2004).

[34] A. Kawai and K. Suzuki, Microelec. Eng. **83**, 655 (2006).

[35] Y. Wang, B. Bhushan, and X. Zhao, Nanotechnology **20**, 045301 (2009).

[36] F. M. Fowkes, In *Contact Angle, Wettability and Adhesion*; Fowkes, F., Ed.; Advances in Chemistry Series, Vol. 43; American Chemical Society: Washington, DC, 1964; Chapter 6.

[37] A. Karagunduz, K. D. Pennell, and M. H. Young, Soil. Sci. Soc. Am. J. **65**, 1392 (2001).

[38] H. Matsubara *et al.*, Coll. Surf. A **301**, 141 (2007).

[39] S. Pegiadou-Koemtzopoulou, I. Eleftheriadis, and A. Kehayoglou, J. Surfact. Deterg. **1**, 59 (1998).

[40] H. Naorem and S. D. Devi, J. Surfac. Sci. Tech. **22**, 89 (2006).

[41] P. Marmottant, S. van der Meer, M. Emmer, M. Versluis, N. de Jong, S. Hilgenfeldt, D. Lohse, J. Acoust. Soc. Am. **118**, 3499 (2005).

[42] J. A. V. Butler, Proc. Roy. Soc. Ser. A **138**, 348 (1932).

[43] V. B. Fainerman and R. Miller, Langmuir **12**, 6011 (1996).

[44] V. B. Fainerman, R. Miller, and R. Wustneck, J. Phys. Chem. B **101**, 6479 (1997).

[45] J. T. Davies, Proc. Roy. Soc. Ser. A **208**, 224 (1951).

[46] J. T. Davies, Proc. Roy. Soc. Ser. A **245**, 417 (1958).

[47] J. T. Davies, Proc. Roy. Soc. Ser. A **245**, 429 (1958).

[48] R. P. Borwankar and D. T. Wasan, Chem. Eng. Sci. **43**, 1323 (1988).

[49] H. Diamant and D. Andelman, J. Chem. Phys. **100**, 13732 (1996).

[50] V. B. Fainerman, Zh. Fiz. Khim. **56**, 2506 (1982).

[51] T. Saitoh, N. Ojima, H. Hoshino, and T. Yotsuyanagi, Mikrochim. Acta **106**, 91 (1992).

[52] V. B. Fainerman and E. H. Lucassen-Reynders, Adv. Coll. Int. Sci. **96**, 295 (2002).

[53] A. Habich, W. A. Ducker, D. E. Dunstan, and X. Zhang, J. Phys. Chem. B **114**, 6962 (2010).

# Figures

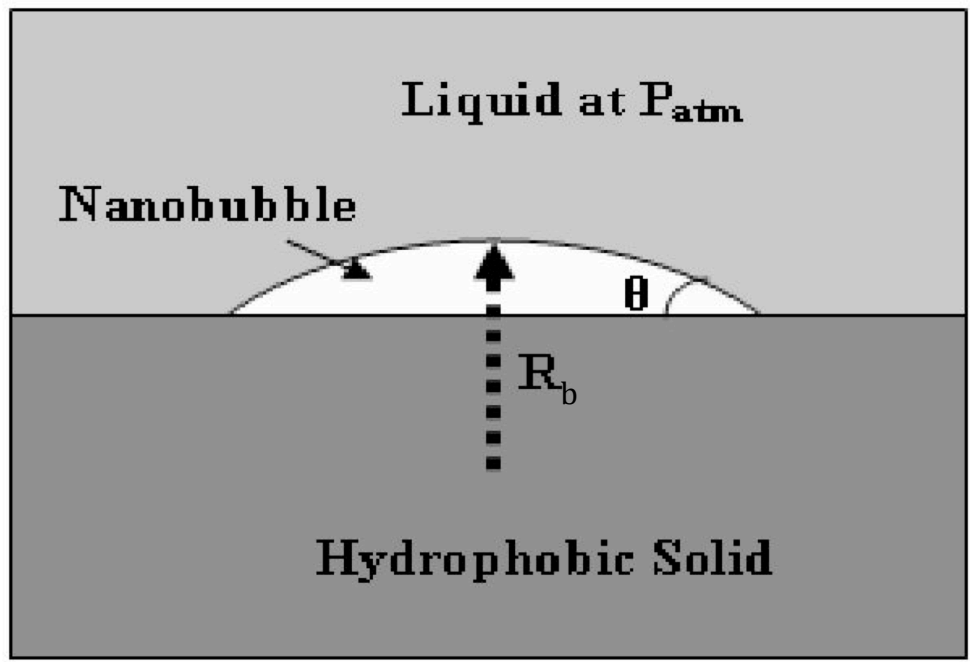

**Figure 1:** Schematic of Nanobubble, with θ being the gas-side contact angle and $R_b$ being the radius of the bubble.

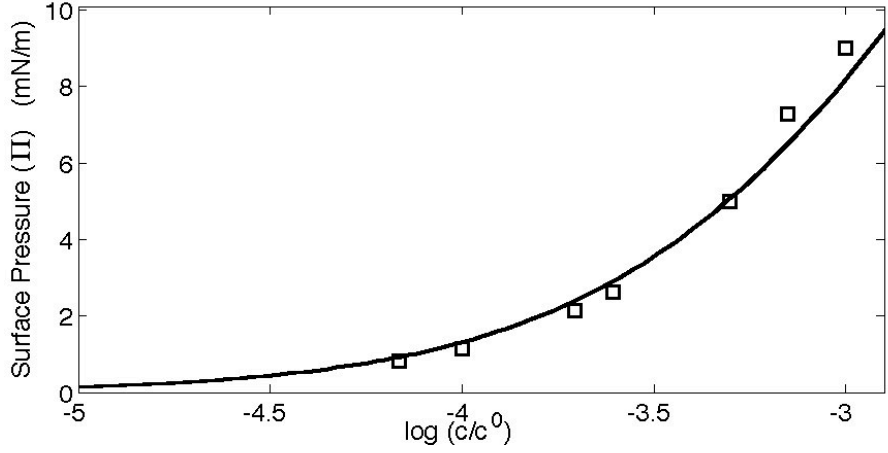

**Figure 2a:** Variation of the surface pressure with bulk ionic concentration (here $c$ is in mol/m$^3$ and $c_0 = 1$ mol/m$^3$) of non-ionic surfactants BHBC$_{16}$. The continuous line is the result obtained from the present simulation (by iteratively solving eqs. 16 and 17, with $\omega_0 = 6.023 \times 10^4$ $m^2/mol$, $\omega_1 = 2.5 \times 10^5$ $m^2/mol$, $K$ (fitting variable) $= 1.9 \times 10^7$, $R = 8.314 J/mol.K$ and $T = 300K$), whereas the squares are the data from experimental results [29,30].

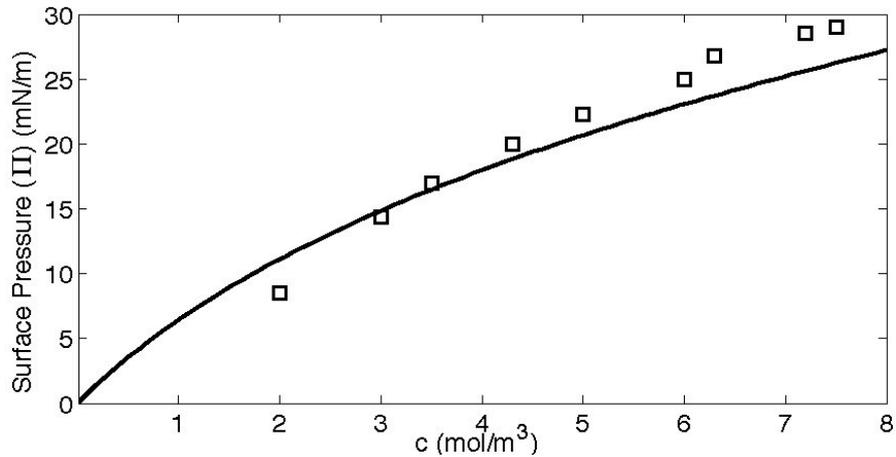

**Figure 2b:** Variation of the surface pressure with bulk ionic concentration (expressed in mol/m$^3$) of ionic surfactants Sodium Dodecyl Sulfate (SDS). The continuous line is the result obtained from the present simulation (by iteratively solving eqs. 16 and 25, with $\omega_0 = 6.023 \times 10^4$ $m^2/mol$, $\omega_1 = 3.4 \times 10^5$ $m^2/mol$ [51], $K$ (fitting variable) $= 4 \times 10^3$, $R = 8.314 J/mol.K$ and $T = 300K$), whereas the squares are the data from experimental results [27,28].

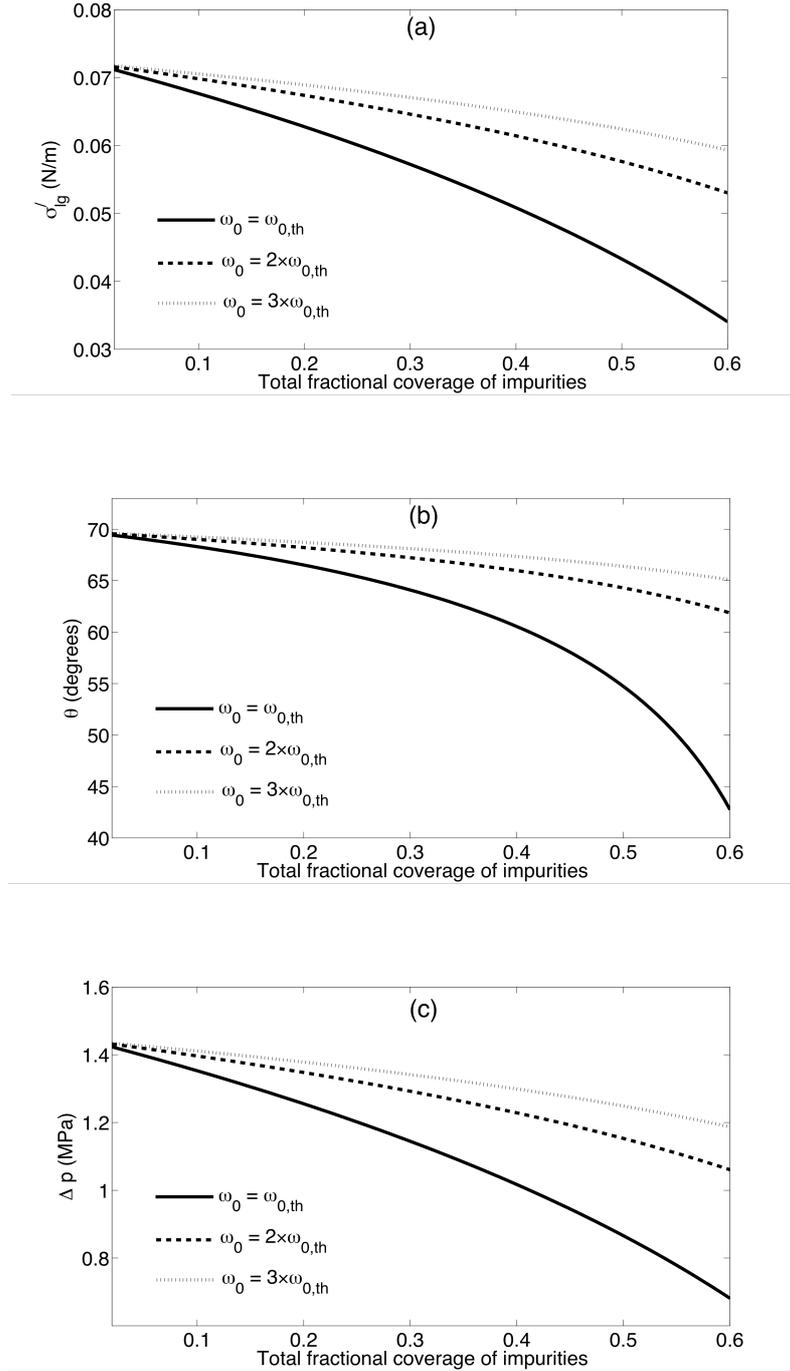

**Figure 3:** Variation of (a) the modified surface tension $\sigma'_{lg}$ (with $\sigma_{lg} = 0.072 N/m$) (b) the gas side contact angle θ (with $\Delta\sigma = \sigma_{sl} - \sigma_{sg} = 0.025 N/m$) and (c) the Laplace pressure $\Delta p$ (with radius of the spherical cap $R_b = 100 nm$) with the total fractional coverage of impurities for different possible values of the partial molar area ($\omega_0$) of the solvent for the case of *non-ionic* surface impurities. In these plots $\omega_{0,th} = 6.023 \times 10^4$ $m^2/mol$. Other constant parameters used for the plots are the gas constant $R = 8.314 J/mol.K$ and $T = 300 K$.

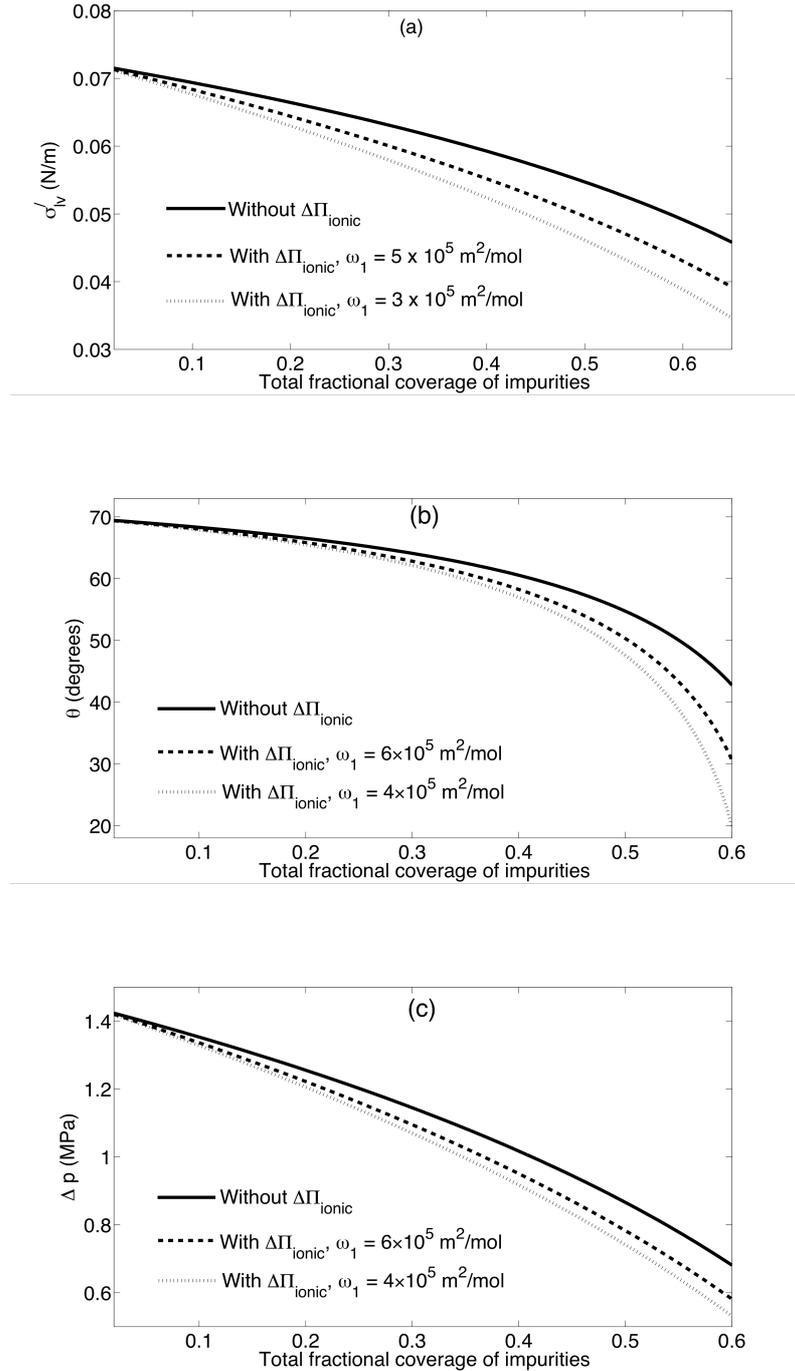

**Figure 4:** Variation of (a) the modified surface tension $\sigma'_{lg}$ (with $\sigma_{lg} = 0.072 N/m$) (b) the gas side contact angle θ (with $\Delta\sigma = \sigma_{sl} - \sigma_{sg} = 0.025 N/m$) and (c) the Laplace pressure $\Delta p$ (with radius of the spherical cap $R_b = 100 nm$) with the total fractional coverage of impurities for the case of *ionic* surface impurities. Results are shown both with and without the ionic contribution to the surface pressure. Here, we study the effects of variation partial molar area of the impurities ($\omega_1$) and consider that there is only one kind of impurity at the interface which is ionic in nature (valence = 1), with all the ions acting as surface active or potential-determining ions. Other parameters used for the plots are $\omega_0 = \omega_{0,th} = 6.023 \times 10^4 \ m^2/mol$, $R = 8.314 J/mol.K$ and $T = 300K$.